\documentclass{aastex631}

\usepackage{multirow}

\usepackage{color}
\usepackage{ulem}
\newcommand{\fra}[1]{\textcolor{blue}{#1}}
\usepackage{comment}

\usepackage{amsmath}
\usepackage{subfigure}
\usepackage{ragged2e}
\shorttitle{AASTeX v6.3.1 Sample article}
\shortauthors{Guastavino et al.}
\graphicspath{{./}{figures/}}
\begin{document}

%\title{video classification for low variance flare forecasting\footnote{Released on March, 1st, 2021}}

% \title{Validation paradigm for supervised machine learning in flare forecasting: an application to video data \footnote{Released on March, 1st, 2021}}
\title{Implementation paradigm for supervised flare forecasting studies: \\ a deep learning application with video data \footnote{Released on March, 1st, 2021}}

\author[0000-0001-7047-1148]{Sabrina Guastavino}
\affiliation{The MIDA group, Dipartimento di Matematica, Università di Genova, \\
Via Dodecaneso, 35, Genova, Italy}

\author[0000-0003-1087-7589]{Francesco Marchetti}
\affiliation{Istituto Nazionale di Alta Matematica (INdAM) - Dipartimento di Matematica \lq\lq Tullio Levi-Civita", Università di Padova, \\
Via Trieste, 63, Padova, Italy}

\author[0000-0002-4776-0256]{Federico Benvenuto}
\affiliation{The MIDA group, Dipartimento di Matematica, , Università di Genova, \\
Via Dodecaneso, 35, Genova, Italy}

\author[0000-0003-2105-8554]{Cristina Campi}
\affiliation{The MIDA group, Dipartimento di Matematica, Università di Genova, \\
Via Dodecaneso, 35, Genova, Italy}

\author[0000-0003-1700-991X]{Michele Piana}
\affiliation{The MIDA group, Dipartimento di Matematica, Università di Genova and CNR - SPIN Genova, \\
Via Dodecaneso, 35, Genova, Italy}
%\affiliation{CNR-SPIN}



\begin{abstract}
Solar flare forecasting can be realized by means of
the analysis of magnetic data through artificial intelligence techniques. The aim is to predict whether a magnetic active region (AR) will originate solar flares above a certain class within a certain amount of time. A crucial issue is concerned with the way the adopted machine learning method is implemented, since forecasting results strongly depend on the criterion with which training, validation, and test sets are populated. In this paper we propose a general paradigm to generate these sets in such a way that they are independent from each other and internally well-balanced in terms of AR flaring effectiveness. This set generation process provides a ground for comparison for the performance assessment of machine learning algorithms. Finally, we use this implementation paradigm in the case of a deep neural network, which takes as input videos of magnetograms recorded by the {\em{Helioseismic and Magnetic Imager}} on-board the {\em{Solar Dynamics Observatory}} ({\em{SDO/HMI}}). To our knowledge, this is the first time that the solar flare forecasting problem is addressed by means of a deep neural network for video classification, which does not require any a priori extraction of features from the {\em{HMI}} magnetograms. %does not need of any a posteriori optimization of skill scores.
\end{abstract}

\keywords{Astronomy data analysis (1858); Neural networks (1933); Solar flares
(1496); Solar activity (1475); Solar active region magnetic fields (1975)}


\section{Introduction} \label{sec:intro}

The problem of solar flare forecasting can be defined as a binary prediction problem, in which the aim is to predict whether, within a certain amount of time, an active region (AR) will originate solar flares of above a given flare class according to the GOES classification \citep{2010hssr.book.....S}. The two main strategies to address this problem rely on either statistical methods \citep{song_statistical_2009, Mason_2010, bloomfield2012toward, Barnes_2016} or machine learning algorithms (see \citep{2021JSWSC..11...39G} and references therein). %\citep{2018ApJ...853...90B,2021JSWSC..11...39G,2020ApJ...904L...7B,2017ApJ...835..156N,2019JSWSC...9A..22F,2018SoPh..293...28F,2018ApJ...856....7H,2020ApJ...890L...5Y,2018ApJ...869...91P}. 

Approaches formulated within the machine learning framework typically need three ingredients: a supervised classification algorithm for classification, a historical data set for its training, and a score for the assessment of performance. However, one of the most intriguing aspects of the flare forecasting game addressed with machine learning is that the studies performed so far lead to significantly different skill scores, although they are applied to the same data archive. Just as a very partial example, Table \ref{table:tab-1} contains the performance outcomes of twelve flare forecasting studies realized by means of machine learning approaches as applied to observations provided by the same space instrument, the {\em{Helioseismic and Magnetic Imager}} on-board the {\em{Solar Dynamics Observatory}} ({\em{SDO/HMI}}). For each one of the twelve studies, the table reports whether a confidence interval on the skill score is computed; whether data belonging to the same AR are split between training and test set; 
%each AR image is present just in either the training or the test set {\bf{(NON MI PIACE MOLTO COME HO SCRITTO QUESTA COSA DEL NON MISCHIARE LE AR)}}; 
whether a validation set is exploited to optimize the machine learning algorithms; which is the machine learning algorithm applied; which are the values obtained for a specific skill score in the case of the prediction of flares with GOES class higher than C (C1+ flares) and M (M1+ flares), respectively. In our opinion, the reason for the heterogeneity of the skill score values in the table is in the fact that there is no general agreement among flarecasters about a validation strategy for the prediction methods. Indeed, given a specific historical archive, these twelve methods generate the training, validation, and test sets according to completely different rules; further, not all methods compute a confidence interval in order to assess the statistical reliability of the scores; and, finally, not all methods distinguish between validation and testing.

The first objective of the present paper is to propose a general paradigm for the implementation and assessment of flare forecasting processes based on supervised machine and deep learning approaches. Specifically, we believe that this kind of strategy should suggest a common perspective about two specific issues: the way data preparation is realized, with specific focus on the way the historical data set is split into a training set, a validation set and a test set; and the way the prediction performances are presented, with specific focus on the computation of the statistical reliability of results. As far as the first issue is concerned, we propose a standardized approach to data splitting based on an accurate definition of data sample, which accounts for the uniformity of training, validation and test sets with respect to both flare classes, and the different possible typologies of null events (this standardized approach should also account for the fact that ARs can be represented by different types of data such as point-in-time vectors or time series of physical features, and {\em{HMI}} images or videos). As far as the second issue is concerned, we point out that any machine learning algorithm should be repeatedly trained on data subsets generated by means of random extractions of ARs from the {\em{HMI}} archive, in order to associate a confidence interval to the skill score values computed on the test set.

The second objective of this paper is to present, for the first time, a deep learning technique that takes as input {\em{HMI}} videos and provides as output a binary prediction with no intermediate processing of the computed features. Our technique is based on a Long-term Recurrent Neural Network (LRCN) architecture \citep{yu2019review}, which combines the use of a Convolutional Neural Network (CNN) for the extraction of morphological features of the ARs together with a Long Short-Term Memory (LSTM) network \citep{hochreiter1997long} for the temporal analysis of the sequences. CNNs for time series of {\em{HMI}} images have been already used by \cite{chen2019identifying}, following an approach that first extracts features from the images by means of an auto-encoder network, then artificially removes redundant features extracted by the CNN according to a $p$-value analysis, and finally organizes the extracted features in time series given as input to an LSTM network computing a binary prediction. 

Differently from the technique used in \cite{chen2019identifying}, our proposed method does not separate the CNN analysis from the LSTM one, and therefore it does not need an a posteriori processing of the features extracted by the CNN. This is a crucial point, since the weights updating process for the autoenconder network in \cite{chen2019identifying}, depends on the optimization of a regression loss, which measures the discrepancy between the reconstructed images and the experimental ones, whereas the weights updating process for the CNN in the LRCN network depends on the optimization of a classification loss, which measures the discrepancy between the predicted probability that an event occurs with the YES-NO actual labels. 

As a final technical detail, in the present study we propose to use a Score Oriented Loss (SOL) function \citep{marchetti2021score} for the network optimization in the training phase, which allows an automated optimization of a given skill score without the need of an a posteriori choice of the optimal threshold that converts the probabilistic outcomes into binary classification. Specifically, the SOL function applied in this paper is based on the optimization of the TSS, which is highly insensitive to the class imbalance ratio in the training set.

The plan of the paper is as follows. Section 2 describes the implementation paradigm. Section 3 focuses on such strategy when applied to video data preparation. Section 4 discusses the design of the applied deep learning model. Section 5 is devoted to the description of the prediction results. Our conclusions are offered in Section 6.

\begin{table}[ht]
		\centering
\label{tab:comparison}
\resizebox{0.9\textwidth}{!}{
\hspace*{-3cm}
\begin{tabular}{|l | l | l | l | l | l | l | l | l | l |}
\hline
\multirow{2}{*}{paper} & \multirow{2}{*}{data} & multiple test & \multirow{2}{*}{AR data split} &  \multirow{2}{*}{validation}  & \multirow{2}{*}{method} & \multirow{2}{*}{score - C1+} & \multirow{2}{*}{score - M1+}\\ 
 & & realizations & & & & \\ 
\hline 
\multirow{2}{*}{\cite{bobra2015solar}} & point in time & \multirow{2}{*}{yes} & \multirow{2}{*}{yes}  & \multirow{2}{*}{yes} & \multirow{2}{*}{SVM} & \multirow{2}{*}{-} & \multirow{2}{*}{$0.74$} \\
& features (SHARP) &  &   &  &  &  &\\
\hline
\multirow{2}{*}{\cite{liu2017predicting}} & point in time & \multirow{2}{*}{yes} & \multirow{2}{*}{yes}  &
\multirow{2}{*}{yes} & \multirow{2}{*}{RF} & \multirow{2}{*}{-}  & \multirow{2}{*}{$0.76$} \\
& features (SHARP) &  &   &  & &  &\\
\hline \multirow{2}{*}{\cite{nishizuka2018deep}} & point in time & \multirow{2}{*}{no} & \multirow{2}{*}{no}  &
\multirow{2}{*}{no} &\multirow{2}{*}{MLP} & \multirow{2}{*}{$0.63$} & \multirow{2}{*}{$0.80$}\\
& features (SHARP + others) &   &   &  & &  &\\
\hline  
\multirow{2}{*}{\cite{florios2018forecasting}} & point in time & \multirow{2}{*}{yes} & \multirow{2}{*}{yes}  &
\multirow{2}{*}{yes} &\multirow{2}{*}{RF} & \multirow{2}{*}{$0.60$} & \multirow{2}{*}{$0.74$} \\
& features (FLARECAST) &  &   &  & &  &\\
\hline
 \multirow{2}{*}{\cite{jonas2018flare}} & point in time & \multirow{2}{*}{yes} & \multirow{2}{*}{no?}  &
 \multirow{2}{*}{yes} & \multirow{2}{*}{RF} & - & \multirow{2}{*}{$0.74-0.81$} \\
& features (SHARP+others) &  &   &  & &  &\\
 \hline
\multirow{2}{*}{\cite{campi2019feature}} & point in time & \multirow{2}{*}{yes} & \multirow{2}{*}{no}  &
\multirow{2}{*}{yes} & \multirow{2}{*}{hybrid lasso} & \multirow{2}{*}{$0.54$} & \multirow{2}{*}{$0.67$} \\
& features (FLARECAST) &  &   &  & &  &\\
  \hline
  
  \multirow{2}{*}{\cite{liu2019predicting}*} & time series & \multirow{2}{*}{yes} & \multirow{2}{*}{no}  &
\multirow{2}{*}{yes}  &  \multirow{2}{*}{LSTM} & \multirow{2}{*}{$0.61$} & \multirow{2}{*}{$0.79$}\\
& features (SHARP) &  &   &  & &  &\\
  \hline
  \multirow{2}{*}{\cite{wang2020predicting}} & time series & .. & \multirow{2}{*}{no}  &
\multirow{2}{*}{yes} & \multirow{2}{*}{LSTM} & \multirow{2}{*}{$0.55$?} & \multirow{2}{*}{$0.68$?}\\
& features (SHARP) &  &   &  & &  &\\
  \hline
%   \multirow{2}{*}{Park et al (2018)} & HMI and & \multirow{2}{*}{Yes} & \multirow{2}{*}{No}  &
% \multirow{2}{*}{Yes} & \multirow{2}{*}{CNN} & \multirow{2}{*}{$0.69$} & ..\\
% & MDI images &  &   &  & &  &\\
 % \hline
   \multirow{2}{*}{ \cite{park2018application}} & HMI and & \multirow{2}{*}{no} & \multirow{2}{*}{no}  &
\multirow{2}{*}{yes} & \multirow{2}{*}{CNN} & \multirow{2}{*}{$0.63$} & .. \\
& MDI images &  &   &  & &  & \\
  \hline
    \multirow{2}{*}{\cite{huang2018deep}} & HMI and & \multirow{2}{*}{y} & \multirow{2}{*}{yes}  &
\multirow{2}{*}{-} & \multirow{2}{*}{CNN} & \multirow{2}{*}{$0.49$} & \multirow{2}{*}{$0.66$} \\
& MDI images &  &   &  & &  & \\
  \hline
  
    \multirow{2}{*}{\cite{li2020predicting}} & HMI & \multirow{2}{*}{yes} & \multirow{2}{*}{no}  &
\multirow{2}{*}{no}  & \multirow{2}{*}{CNN} & \multirow{2}{*}{$0.68$} & \multirow{2}{*}{$0.75$}\\
& images &  &   &  & &  & \\
  \hline
  
   \multirow{2}{*}{\cite{yi2021visual}} & HMI & \multirow{2}{*}{no} & \multirow{2}{*}{no}  &
\multirow{2}{*}{yes} & \multirow{2}{*}{CNN} & \multirow{2}{*}{$0.65$} & - \\
& images &  &   &  & &  &\\
  \hline
\end{tabular}
}
\caption{Description of twelve flare forecasting studies based on machine learning. For each study, the table reports: the main author (column "paper"); the kind of data used (column "data"); whether a confidence strip has been computed for the skill score (column "multiple test realizations"); whether data belonging to the same AR are split between the training and test sets (column "AR data split"); whether a validation set has been used to optimize the machine learning algorithm (column "validation"); which method has been used (column "method"); the score values for the prediction of C1+ and M1+ flares, respectively (columns "score - C1+" and "score - M1+", respectively).  }\label{table:tab-1}
\end{table}

%Aggiungere in tabella il ratio tra positivi e negativi (o percentuale esempi positivi) e la percentuale di esempi positivi con flares > M (quando possono essere calcolati).
% Traditional machine-learning methods commonly used for flare
% prediction cover artificial neural network (Qahwaji &
% Colak 2007; Ahmed et al. 2013; Li & Zhu 2013; Nishizuka
% et al. 2018), k-nearest neighbors (Li et al. 2008; Huang
% et al. 2013), support vector machine (Yuan et al. 2010;
% Bobra & Couvidat 2015; Nishizuka et al. 2017; Sadykov &
% Kosovichev 2017), random forests (Liu et al. 2017; Florios
% et al. 2018), and ensemble learning (Colak & Qahwaji 2009;
% Huang et al. 2010; Guerra et al. 2015). 

\section{Implementation and assessment paradigm}\label{sec:val_strat}

\subsection{Sample definition}\label{sec:generation of sets}
Given an AR, we split the data associated with it into contiguous {\it samples}, each one corresponding to a time interval of fixed duration.
When the interval is reduced to only one time point, each sample can be a set of numerical values, for example values of physical features of that AR, or an AR image. Alternatively, when the time interval is bigger than one time point, a sample can be a time series of features, or a video of magnetograms.
% If we divide each active region according to time intervals of fixed duration we obtain that the active region is composed of samples of the same time length.
% The samples in question can have dimension 0, for example a feature or a set of features, or 1d, for example a feature for each moment in time, or 2d, for example images if the interval is reduced to only 1 time interval, or even 3d videos.
Each data sample has then been labelled with a mark of type X, M, C, NO1, NO2, NO3, NO4
depending on the ability of the AR to which the sample belongs either to generate or not to generate a flare:
\begin{itemize}
\item X class sample: sample of AR that originated a flare in the next 24 hours after the sample time, with maximum flare class X1 or above.
\item M class sample: sample of AR that originated a flare in the next 24 hours after the sample time, with maximum flare class M1 or above but under class X.
\item C class sample: sample of AR that originated a flare in the next 24 hours after the sample time, with maximum flare class C1 or above but under class M.
\item NO1 class sample: sample of AR that never originated a C1+ flare.
\item NO2 class sample: sample of AR that did not originate a C1+ flare in the next 24 hours after the sample time, it did not originate a C1+ flare in the past, but it did originate a C1+ flare in the future.
\item NO3 class sample: sample of AR that did not originate a C1+ flare in the next 24 hours after the sample time, but it did originate a C1+ flare in the 48 hours before the sample time.
\item NO4 class sample: sample of AR that did not originate a C1+ flare in the next 24 hours after the sample time and did not originate a C1+ flare in the 48 hours before the sample time, but it did originate a C1+ flare before the 48 hours before the sample time.
\end{itemize}

We can therefore think of an AR as a set of data samples labeled according to the previous criterion. For example, suppose that the AR number $12645$ includes $4$ X samples, $10$ M samples, $24$ C samples and $13$ NO2 samples; then this AR can be described by means of the notation: $\mathrm{AR}_{12645} =\{4\mathrm{X}, 10\mathrm{M}, 24\mathrm{C}, 13\mathrm{NO2}  \}$ .

\subsection{Well-balanced data sets}\label{sec:balan}

The procedure for the generation of training, validation and test sets was based on two criteria.
\begin{enumerate}
\item {\bf{Proportionality}}. We required the sets to have almost equal rates of samples for each sample type described above. In order to construct sets as reliable as possible, we required the rates to be similar to the ones characterizing the historical archive; i.e., in our experiments we set the following rates, coherent to the ones in the {\em{HMI}} archive for the time interval between 2012, September 14 and 2017, September 30 (where $p_X$ denotes the rate of the X class sample, $p_M$ that of the M class sample, and so on):
\begin{itemize}
\item $p_\mathrm{X} \approx 0.13\%$
\item $p_\mathrm{M} \approx 3.21\%$
\item $p_\mathrm{C} \approx 18.08\%$
\item $p_{\mathrm{NO1}} \approx 45.94\%$
\item $p_{\mathrm{NO2}} \approx 3.57\%$
\item $p_{\mathrm{NO3}} \approx 12.06\%$
\item $p_{\mathrm{NO4}} \approx 17.01\%$
\end{itemize}
\item {\bf{Parsimony}}. We wanted that each subset of samples came from as few ARs as possible.
In this way, we promoted training, validation and test sets to be independent from each other, in the sense that samples belonging to the same AR must fall into the same data set. %Let us consider an entire dataset of active regions. 
%Say $n$ the number of samples composing the AR dataset.
%The proportion between the samples of type X, M, C, No$_{after}$, No$_{before}$, No$_{middle}$, No$_{never}$ is given. 
%There are $n_X$ samples labelled with $X$, $n_M$ samples labelled with $M$ and so on.
%We have $n_i = n * p_i$ with $i=$X, M, C, No$_{after}$, No$_{before}$, No$_{middle}$, No$_{never}$ and in our case 
\end{enumerate}

\subsection{Procedure for data sets generation}\label{sec:gener}
We first set the size of the data set we wanted to create equal to $n$. This implied that we needed $n_\mathrm{X} = n \cdot p_\mathrm{X}$ samples labelled with X, $n_\mathrm{M} = n \cdot p_\mathrm{M}$ samples labelled with M, and so on. The procedure is as follows:
\begin{enumerate}
\item We randomly took an AR containing X flares and put all the samples of this AR in our data set. We kept on including new ARs with X flares, until the number of samples labelled with X became $n_\mathrm{X}$. 
If an AR contained more X flares than needed, we discarded those in excess. 
\item We checked the amount of M flares in the data set under construction. If we had more than $n_\mathrm{M}$ samples with M flares, we randomly discarded those in excess; in the case the number was smaller, we randomly included ARs containing M flares (but not X flares) up to the correct rate. 
\item The process continued until we had the prescribed number of samples for each type.
\end{enumerate}
In order to collect NO1, NO2, NO3, and NO4 data samples, we used just ARs not containing X and M flares, since these latter ones are precious for the construction of other independent and well balanced data sets according to the parsimony and proportionality criterion, respectively.

We point out that the obtained algorithm is sub-optimal, since it may not find the smallest number of ARs needed, but it allows constructing a wide variety of independent data sets as it operates randomly.

\subsection{Algorithm validation}

Each machine learning algorithm depends on some parameters (e.g. weights utilized by each neuron of the neural network), which must be optimised on the basis of the historical data set. To this aim, the performances of the algorithm have been evaluated on a validation set and the weights’ values corresponding to the best validation score are employed in the test phase.

\subsection{Assessment of results}\label{sec:comp_results}
In order to compare the performances of different machine learning methods in flare forecasting, the following items should be accounted for:

\begin{enumerate}
    
    \item
    
    % The classification results should be evaluated by considering skill scores that are suitable for imbalanced data classification. Indeed, solar events are relatively \textit{rare}, as already pointed out in Section \ref{sec:generation of sets}. Therefore, a chosen score needs to be capable of representing the performance of the classifier concerning the \textit{small} positive class.
    
    % Examples of widely-adopted scores for flare forecasting applications are the True Skill Statistic (TSS) \cite{hanssen1965relationship} or the Heidke Skill Score (HSS) \cite{Heidke1926}.
     The classification results should be evaluated by considering appropriate skill scores defined on the so-called confusion matrix, which is characterized by four elements: true positives (TPs), i.e. the number of samples labeled with YES and correctly predicted as positive; true negatives (TNs), i.e. the number of samples labeled with NO and correctly predicted as negative; false positives (FPs), i.e. the number of samples labeled with NO incorrectly predicted as positive; and false negatives (FNs), i.e. the number of samples labeled with YES and incorrectly predicted as negative. In solar flare forecasting the most meaningful skill scores are the ones specific for imbalanced data classification. Indeed, solar events are relatively seldom, as already pointed out in Section \ref{sec:generation of sets}. Therefore, a chosen score needs to be able to represent the performance of the classifier concerning data sets with a small number of positive events.
     Among all possible skill scores, the True Skill Statistic (TSS) \citep{hanssen1965relationship} is defined as \begin{equation}
    \mathrm{TSS}=\frac{\mathrm{TP}}{\mathrm{TP}+\mathrm{FN}}-\frac{\mathrm{FP}}{\mathrm{FP}+\mathrm{TN}}=\mathrm{POD}-\mathrm{FAR}~,
\end{equation}
and its values have range in the interval $[-1,1]$: when $\mathrm{TSS}=1$, the performance is optimal, while $\mathrm{TSS}>0$ means that the rates of positive and negative events are mixed up. The TSS is insensible to the class-imbalance ratio \citep{bloomfield2012toward}, and therefore this is the skill score adopted in the present study.

    \item    The strategy outlined throughout Sections \ref{sec:generation of sets}, \ref{sec:balan}, \ref{sec:gener} ought to be repeated several times in order to achieve some statistical significance. Therefore, many classification tests should be carried out by generating different triples of training, validation and test sets by randomly extracting AR images from the {\em{HMI}} archive.
\end{enumerate}
    
    Once the results are obtained, some statistical indicators such as the mean, standard deviation, maximum and minimum value should be reported. Obviously, the results achieved on the test set should not be produced by applying any validation procedure directly on the test set.

\section{Video data preparation}

In general, the archive of the {\em{SDO/HMI}} mission includes 2D images of continuous intensity, of the full three-component magnetic field vector, and of the line-of-sight magnetic intensity. In the present study, we considered the Near Realtime Space Weather {\em{HMI}} Archive Patch (SHARP) data products associated to the line-of-sight components in the time range between 2012, September 14 and 2017, September 30. More specifically, our data products were $24$ hour long videos made of $40$ SHARP images of an AR, with $36$ minutes cadence. Each image in these time series have been resized to a $128 \times 128$ pixels dimension, due to computational reasons.

\begin{figure}[h!]
    \centering
     \subfigure[{X class video}]{\includegraphics[width=0.8\textwidth]{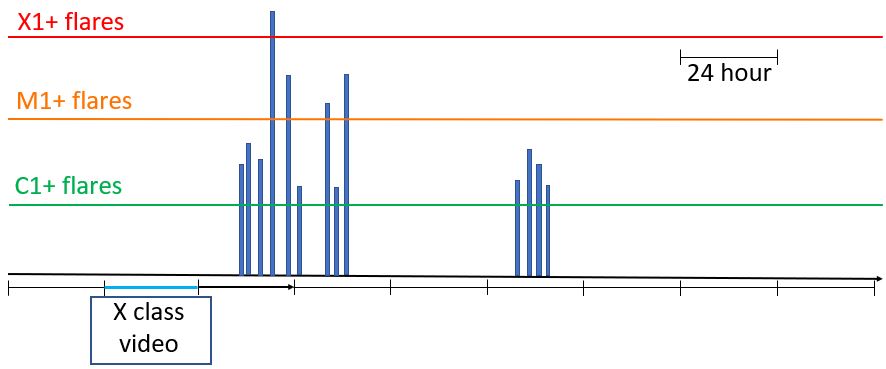}}\\
    \subfigure[{M class video}]{\includegraphics[width=0.8\textwidth]{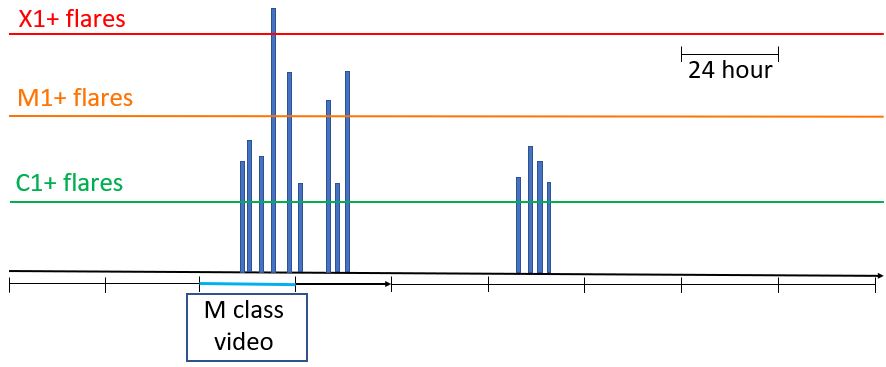}}\\
     \subfigure[{C class video}]{\includegraphics[width=0.8\textwidth]{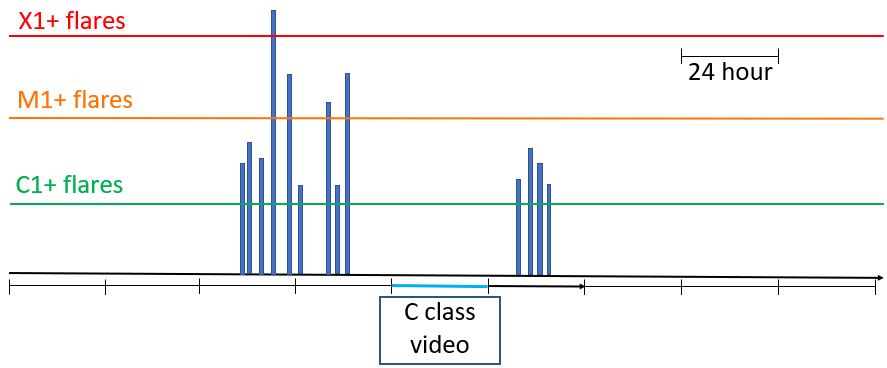}}\\
     \caption{From top to bottom: examples of X class, M class and C class video.}
    \label{fig:X_M_C_class_videos}
\end{figure}

\begin{figure}[h!]
    \centering
    \subfigure[{NO2 video}]{\includegraphics[width=0.8\textwidth]{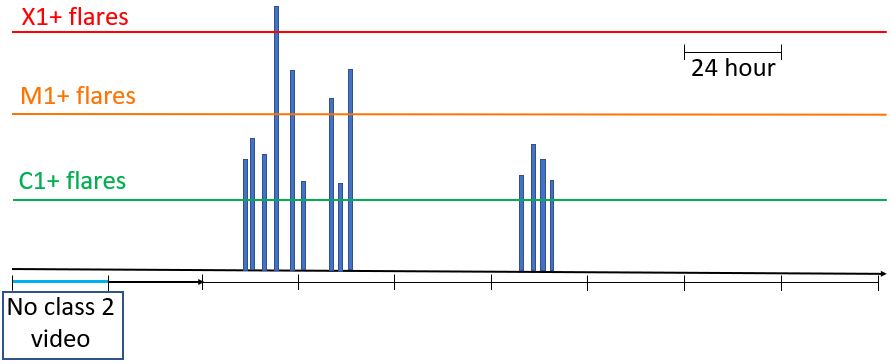}}\\
    \subfigure[{NO3 video}]{\includegraphics[width=0.8\textwidth]{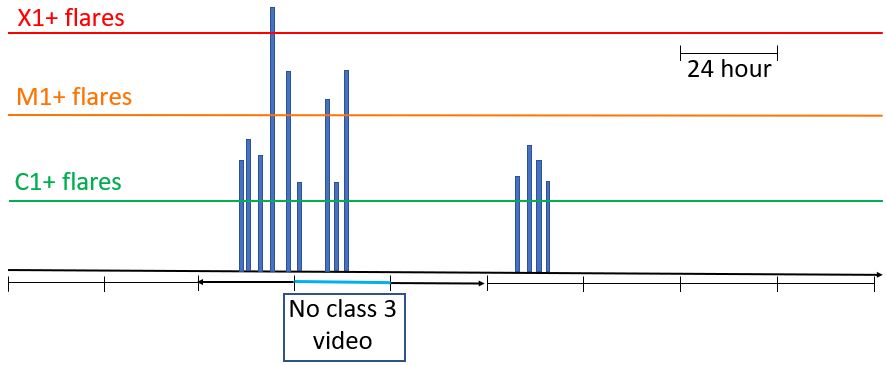}}\\
    \subfigure[{NO4 video}]{\includegraphics[width=0.8\textwidth]{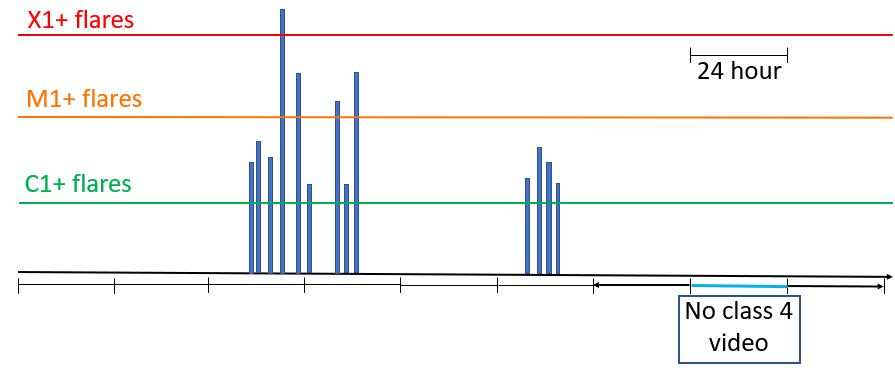}}
     \caption{From top to bottom: examples of NO2, NO3 and NO4 class video.}
    \label{fig:No_class_videos}
\end{figure}

Figure \ref{fig:X_M_C_class_videos} and Figure \ref{fig:No_class_videos} are iconographical representations of how these videos are categorized with respect to the definitions given in Section 2. In particular, on the one hand Figure \ref{fig:X_M_C_class_videos} illustrates a typical temporal history (from its birth to its death) of an AR that originates X1+, M1+, and C1+ flares (we point out that when the aim is to predict C1+ flares then the three kinds of videos are labeled with $1$, whereas when the aim is to predict M1+ flares then just the first two kinds of videos are labeled with $1$). On the other hand, Figure \ref{fig:No_class_videos} provides schematic examples of NO2, NO3, and NO4 videos. NO1 data samples are not included in the figure, since they correspond to ARs that never originated a flaring event.

%\subsection{Data-splitting method}\label{sec:data-splitting}

\section{Deep learning method for HMI video}
\begin{figure}[h!]
    \centering
     \subfigure[{LRCN architecture}]{\includegraphics[width=0.8\textwidth]{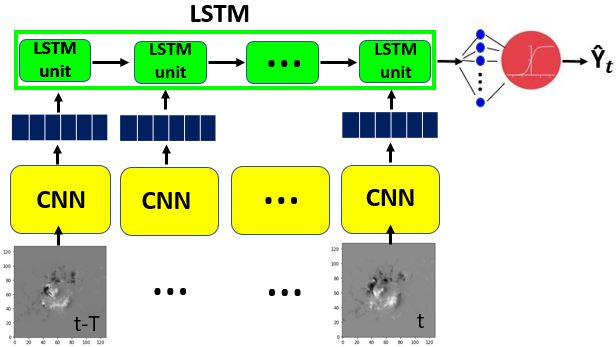}}\\
    \subfigure[{CNN architecture}]{\includegraphics[width=0.6\textwidth]{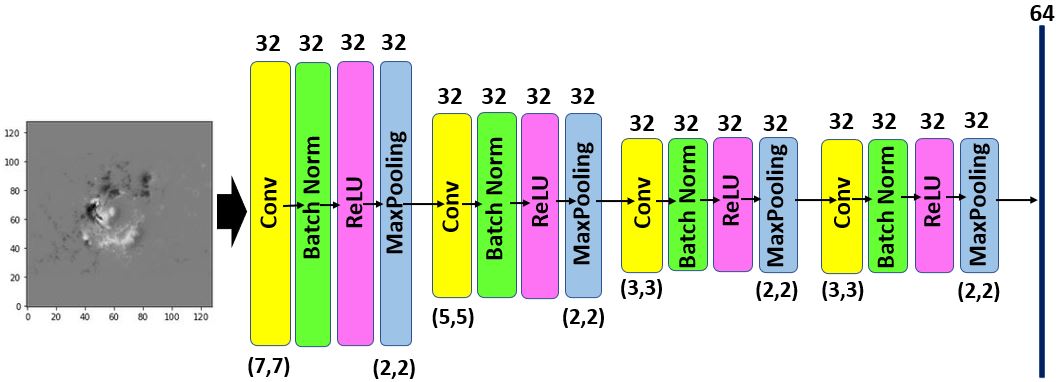}}\\
     \caption{Top panel: The overall LRCN design. Bottom panel: the CNN architecture.}
    \label{fig:deep_model}
\end{figure}

%\subsection{Architecture}
The analysis of the video data samples has been performed by means of a Long-term Recurrent Convolutional Network (LRCN), which is a mixed deep learning model made of a convolutional neural network (CNN) and a Long-Short Term Memory (LSTM) network. The first part of the LRCN network is made of the following sequence of layers (see Figure \ref{fig:deep_model}, top panel): 
\begin{itemize}
    \item
    a $7\times 7$ convolutional layer of $32$ units; a $2\times 2$ max-pooling layer;
    \item 
    a $5\times 5$ convolutional layer of $32$ units; a $2\times 2$ max-pooling layer;
    \item
    a $3\times 3$ convolutional layer of $32$ units; a $2\times 2$ max-pooling layer;
    \item 
    a $3\times 3$ convolutional layer of $32$ units; a $2\times 2$ max-pooling layer; 
    \item a dense layer of $64$ units, where \textit{dropout} is applied with a fraction of $0.1$ input units dropped.
\end{itemize}
Height and width strides are set to $2$ for the convolutional layers and to $1$ for the max-pooling. Each convolutional layer is $L_2$-regularized and the corresponding output is standardized. Before applying the dense layer, the last pooling layer is flattened. The Rectified Linear Unit (ReLU) is used as activation function in all layers. We also point out that the input videos, which consist of $40$ frames of $128\times 128$ images each, are treated as time series, so that the CNN architecture described above is applied to each video frame in parallel.

In the second part of the LRCN, the outputs of the CNNs ($40$ vectors, each one composed by $64$ features) are sequentially considered in time and then passed to the LSTM (see \ref{fig:deep_model}, bottom panel), which consists of $50$ units. Similarly to the dense layer, here dropout is applied with a fraction of $0.5$ active units. Finally, a dense sigmoid unit drives the output of the LSTM to be in the interval $[0,1]$, in order to perform binary classification.
The CNN-LSTM network is trained for 100 epochs by taking batches of $128$ samples. Moreover, the \textit{Adam} scheme \citep{Kingma14} is adopted for the optimization of the weights.

%As far as the loss function utilized in the training phase is concerned, we have adopted the score-driven strategy illustrated in \cite{}. The construction of score-oriented loss (SOL) functions relies on the definition of the probabilistic confusion matrix
%\begin{equation}\label{confusion-matrix}
%\mathrm{CM}(\tau)= 
%\begin{pmatrix}
%\mathrm{TN}(\tau) & \mathrm{FP}(\tau) \\
%\mathrm{FN}(\tau) & \mathrm{TP}(\tau)
%\end{pmatrix},
%\end{equation}
%in which the threshold parameter $\tau \in (0,1)$ is dealt with as a random variable associated to a specific probability density function. From this matrix it is possible to construct, for example, the probabilistic version of the TSS
%\begin{equation}\label{eq:tss_def}
%\textrm{TSS}(\tau)=\frac{\mathrm{TP}(\tau)}{\mathrm{TP}(\tau)+\mathrm{FN}(\tau)}-\frac{\mathrm{FP}(\tau)}{\mathrm{FP}(\tau)+\mathrm{TN}(\tau)}-1 \ ,
%\end{equation}
%and from this the probabilistic TSS-driven loss function
%\begin{equation}\label{TTS-loss}
%l_{TSS} := - \textrm{TSS}(\tau)\ .
%\end{equation}
%This function is differentiable and therefore can be easily minimized in the training phase, with the crucial advantage that the corresponding skill score is automatically optimized, without the need of any a posteriori tuning of the thresholding parameter $\tau$, which is set to the default value $0.5$.

As far as the loss function utilized in the training phase is concerned, we have adopted the score-driven strategy illustrated in \cite{marchetti2021score}. The classical confusion matrix depends on a fixed threshold parameter $\tau \in (0,1)$, i.e., 
\begin{equation}\label{confusion-matrix}
\mathrm{CM}(\tau)= 
\begin{pmatrix}
\mathrm{TN}(\tau) & \mathrm{FP}(\tau) \\
\mathrm{FN}(\tau) & \mathrm{TP}(\tau)
\end{pmatrix}.
\end{equation}
For the construction of score-oriented loss (SOL) functions, the threshold parameter $\tau$ is dealt with as a random variable associated to a specific probability density function. Letting $\mathbb{E}_{\tau}[\cdot]$ be the expected value with respect to $\tau$, we took an expected confusion matrix
\begin{equation}\label{confusion-matrix}
\mathbb{E}_{\tau}[\mathrm{CM}(\tau)]= 
\begin{pmatrix}
\mathbb{E}_{\tau}[\mathrm{TN}(\tau)] & \mathbb{E}_{\tau}[\mathrm{FP}(\tau)] \\
\mathbb{E}_{\tau}[\mathrm{FN}(\tau)] & \mathbb{E}_{\tau}[\mathrm{TP}(\tau)]
\end{pmatrix}.
\end{equation}
From this matrix it was possible to construct the expected TSS
\begin{equation}\label{eq:tss_def}
\mathbb{E}_{\tau}[\textrm{TSS}(\tau)]=\frac{\mathbb{E}_{\tau}[\mathrm{TP}(\tau)]}{\mathbb{E}_{\tau}[\mathrm{TP}(\tau)+\mathrm{FN}(\tau)]}-\frac{\mathbb{E}_{\tau}[\mathrm{FP}(\tau)]}{\mathbb{E}_{\tau}[\mathrm{FP}(\tau)+\mathrm{TN}(\tau)]}-1 \ ,
\end{equation}
and from this the TSS-driven loss function
\begin{equation}\label{TTS-loss}
\ell_{\mathrm{TSS}} := - \mathbb{E}_{\tau}[\textrm{TSS}(\tau)]\ .
\end{equation}
This function is differentiable and therefore can be easily minimized in the training phase, with the crucial advantage that the corresponding skill score is automatically optimized, without the need of any a posteriori tuning of the thresholding parameter $\tau$, which is set to the default value $0.5$.

\section{Results}
The LRCN described in the previous section has been applied to video data generated as described in Section 3, using the validation strategy illustrated in Section 2. We first filled up a training set, a validation set, and a test set made of $3000$, $750$, and $750$ data samples, respectively, and we used data augmentation to increase the cardinality of these sets up to $15000$, $3750$, and $3750$, respectively (the data augmentation process here relied on image rotation and reflection). We repeated this set generation process ten times in order to create ten random realizations of this triple of sets. Table \ref{tab:tss_val_test} shows the prediction results obtained by the LRCN in the case of the realization of the validation and test sets: specifically, the table focuses on TSS, and provides its mean and standard deviation values, the minimum value, the 25th percentile value, the median value, the 75th percentile value and the maximum value for the prediction of both C1+ and M1+ flares. The numbers in this table show that image-based deep learning is more effective in predicting M1+ flares than C1+ flares, in line with most results in the scientific literature. Further, both the mean and the median values for the test sets are rather close to the ones provided by the network when applied to the validation sets, and in all cases the standard deviations are nicely small. 

\begin{figure}[h!]
    \centering
     \subfigure[{C1+ flares prediction}]{\includegraphics[width=0.4\textwidth]{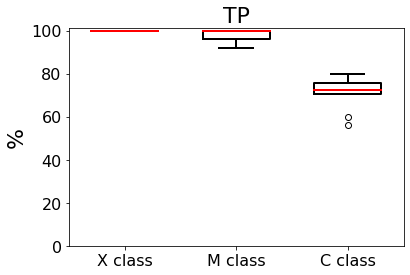}}
    \subfigure[{M1+ flares prediction}]{\includegraphics[width=0.4\textwidth]{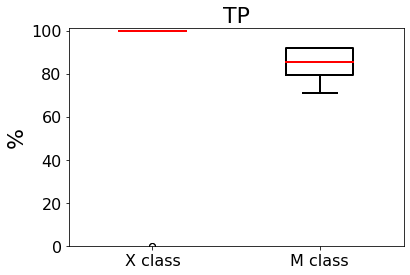}}\\
     \subfigure[{C1+ flares prediction}]{\includegraphics[width=0.4\textwidth]{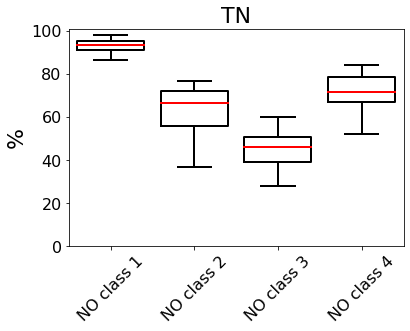}}
    \subfigure[{M1+ flares prediction}]{\includegraphics[width=0.4\textwidth]{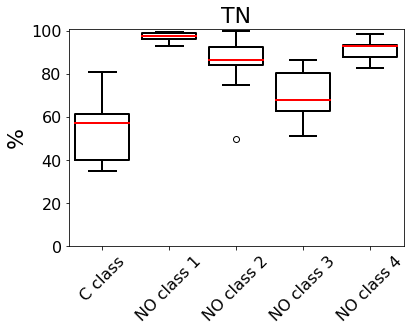}}
     \caption{Percentages of correctly predicted as TP (top panels) and as TN (bottom panels) computed on the $10$ test set for the C1+ flares prediction (first) and for the M1+ flares prediction (second column) with respect to each kind of samples as shown in the x-axis.}
    \label{fig:TP_TN}
\end{figure}

\begin{figure}[h!]
    \centering
     \subfigure[{C1+ flares prediction}]{\includegraphics[width=0.8\textwidth]{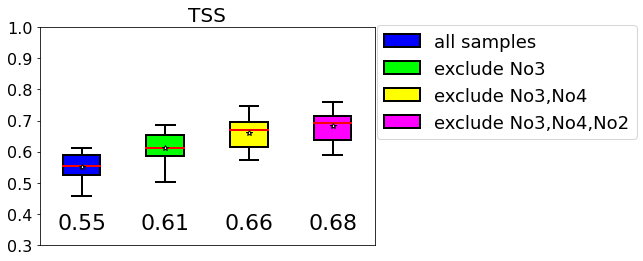}}\\
    \subfigure[{M1+ flares prediction}]{\includegraphics[width=0.8\textwidth]{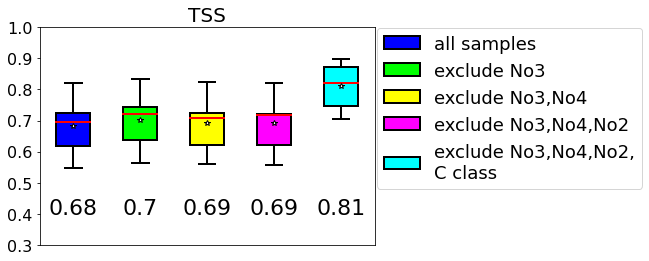}}
     \caption{Boxplots of TSS computed on $10$ test sets for the prediction of C1+ class flares (top panel) and M1+ class flares (bottom panel) by excluding each time a kind of No labeled videos. Top panel. From left to right: (1) TSS computed on overall samples; (2) TSS computed on samples by excluding the NO3 videos; (3) TSS computed on samples by excluding the NO3 and NO4 videos;
   (4) TSS computed on samples by excluding the NO3, NO4 and NO2 videos.
   Bottom panel. From left to right: (1) TSS computed on overall samples; (2) TSS computed on samples by excluding the NO3 videos; (3) TSS computed on samples by excluding the NO3 and NO4 videos;
   (4) TSS computed on samples by excluding the NO3, NO4 and NO2 videos;
   (5) TSS computed on samples by excluding the NO3, NO4, NO2 and C class videos.}
    \label{fig:excluding_No}
\end{figure}

Figure \ref{fig:TP_TN} and Figure \ref{fig:excluding_No} aim at providing a quantitative confirmation of the implementation paradigm introduced in Section 2. The boxplots in Figure \ref{fig:TP_TN} represent the rates of TPs and TNs computed on the ten random realizations of the test set. They show that:
\begin{itemize}
    \item When the aim is to predict C1+ flares, X class and M class flares are predicted with a higher success rate with respect to C flares, coherently to the fact that ARs labelled as either M or X are more distinguishable from ARs associated to NO-class flares (this is particularly true for all cases when flares are close to strong B events).
    \item The prediction of null events does not have the same success rate for all NO-class flares. Indeed, the rate of TNs is significantly higher when the LRCN aims to predict NO1 events, coherently to the fact that data sampled belonging to the other three NO-classes are similar to C1+ data samples (this is particularly true for NO3 data samples, whose prediction has indeed the lowest success rate).
    \item When the aim is to predict M1+ flares, the biggest number of TNs refer to C class data samples, coherently to the fact that such videos may be associated to events with energy budget close to the ones of weak M flares.
\end{itemize}

In Figure \ref{fig:excluding_No} we computed the TSS values while successively excluding data samples of different classes from the test sets. We found out that, in the case of predicting C1+ events, the TSS values are the smallest ones when all classes are represented in the test sets, and nicely increases while successively and cumulatively excluding NO2, NO3, and NO4 events. On the other hand, while predicting M1+ flares, the TSS values significantly increase when data samples belonging to all classes but class M1+ are excluded from the test sets.

\begin{center}
\begin{table}[ht]
\centering
\caption{Mean, standard deviation, minimum, $25$th percentile, median, $75$th percentile and maximum value of the TSS distribution computed on $10$ validation sets and $10$ test sets, separately. \vspace*{0.5cm}
}
\label{tab:tss_val_test}
\resizebox{0.8\textwidth}{!}{
\hspace*{-2.5cm}
\begin{tabular}{|l | c c c c c c c |}
\hline
 & \multicolumn{7}{c|}{TSS (C1+ flares)} \\
 \hline
  &  Mean &  Std &   Min & $25$th perc & Median & $75$th perc & Max  \\
\cline{2-8}
Validation & $0.57$ & $0.02$ & $0.55$ & $0.56$ & $0.57$ & $0.59$ & $0.61$ \\
Test & $0.55$ & $0.05$ & $0.46$ & $0.52$ & $0.54$ &  $0.60$ & $0.61$ \\
\hline
 & \multicolumn{7}{c|}{TSS (M1+ flares)} \\
 \hline
  &  Mean &  Std &   Min & $25$th perc & Median & $75$th perc & Max  \\
  \cline{2-8} 
Validation & $0.76$ & $0.07$ & $0.65$ & $0.67$ & $0.77$ & $0.82$ & $0.85$ \\
Test & $0.68$ & $0.09$ & $0.55$ & $0.61$ & $0.69$ & $0.72$ & $0.82$ \\
\hline
\end{tabular}
}
\end{table}
\end{center}

\section{Comments and conclusions}
The scientific rationale of the present study was two-fold. On the one hand, we aimed at verifying the feasibility of a fully automated flare forecasting procedure that takes as input videos of line-of-sight magnetograms and provides binary predictions as output. On the other hand, we aimed at studying the impact of the data set preparation on the deep learning performances. The conclusions we can draw from this analysis are that, first of all, deep learning is able to realize flare videos classification with prediction performances that are in line with the ones obtained by machine learning approaches that require an a priori extraction of features from the {\em{HMI}} images by means of pattern recognition and feature computation algorithms. Further, in our implementation of the convolutional network, the use of a SOL function allows an a priori optimization of the TSS, which allows avoiding the application of (often critical) a posteriori thresholding on the score value.

Second, the results in Figure \ref{fig:TP_TN} and in Figure \ref{fig:excluding_No} clearly show that the way the training and validation sets are prepared for the network optimization has a really significant impact on the prediction performances \citep{campi2019feature}. In particular, these figures prove that an appropriate balancing of these sets should account for not only the presence of ARs generating flares but even the presence of ARs associated to null events of different kinds.

We finally point out the that TSS values obtained by this analysis are distinctively different from $1$, as typically occurs in most flare forecasting studies based on machine learning. This seems to confirm once more \citep{campi2019feature} the presence of an intrinsically stochastic component of the flaring process, which implies that the flare forecasting challenge has still a predominant probabilistic aspect. In this perspective, two possible futher research lines are the combination of image-based features with features that rely on non-iconographic information within fully data-driven models; and the exploitation of physical information in the design and optimization of the networks.

\bibliography{sample631}{}
\bibliographystyle{aasjournal}



\end{document}